\documentclass[5p,times,twocolumn]{elsarticle}
\usepackage{lineno,hyperref}
\usepackage{subcaption}
\usepackage{tikz}
\usepackage{pgfplotstable}
\usepackage{pgfplots}
\usepackage{multirow}
\usepackage{float}
\usepackage{textcomp}
\usepackage{amsmath,amsfonts,amssymb,bbm}	
\usepackage{multicol}
\usepackage{cuted}
\usepackage{makecell}
\usepackage{colortbl}

\usepackage{booktabs}
\usepackage{xcolor,soul}

\usepackage[section]{placeins}

\modulolinenumbers[5]

\begin{document}

\begin{frontmatter}

\title{SPOCKMIP: Segmentation of Vessels in MRAs with Enhanced Continuity using Maximum Intensity Projection as Loss}

\author[1]{Chethan Radhakrishna}
\author[1]{Karthikesh Varma Chintalapati}
\author[1]{Sri Chandana Hudukula Ram Kumar}
\author[1]{Raviteja Sutrave}


\author[2]{Hendrik Mattern} 
\author[2,4,5]{Oliver Speck}
\author[1,3,5]{Andreas Nürnberger}
\author[1,3,6]{Soumick Chatterjee\corref{mycorrespondingauthor}}
\cortext[mycorrespondingauthor]{Corresponding author:}
\ead{contact@soumick.com}


\address[1]{Faculty of Computer Science, Otto von Guericke University Magdeburg, Germany}
\address[2]{Biomedical Magnetic Resonance, Otto von Guericke University Magdeburg, Germany}
\address[3]{Data and Knowledge Engineering Group, Otto von Guericke University, Magdeburg, Germany}
\address[4]{German Centre for Neurodegenerative Disease, Magdeburg, Germany}
\address[5]{Centre for Behavioural Brain Sciences, Magdeburg, Germany}
\address[6]{Genomics Research Centre, Human Technopole, Milan, Italy}

\begin{abstract}
Identification of vessel structures of different sizes in biomedical images is crucial in the diagnosis of many neurodegenerative diseases. However, the sparsity of good-quality annotations of such images makes the task of vessel segmentation challenging. Deep learning offers an efficient way to segment vessels of different sizes by learning their high-level feature representations and the spatial continuity of such features across dimensions. Semi-supervised patch-based approaches have been effective in identifying small vessels of one to two voxels in diameter. This study focuses on improving the segmentation quality by considering the spatial correlation of the features using the Maximum Intensity Projection~(MIP) as an additional loss criterion. Two methods are proposed with the incorporation of MIPs of label segmentation on the single~(z-axis) and multiple perceivable axes of the 3D volume. The proposed MIP-based methods produce segmentations with improved vessel continuity, which is evident in visual examinations of ROIs. In this study, a UNet MSS with ReLU activation replaced by LeakyReLU is trained on the Study Forrest dataset. Patch-based training is improved by introducing an additional loss term, MIP loss, to penalise the predicted discontinuity of vessels. A training set of 14 volumes is selected from the StudyForrest dataset comprising of 18 7-Tesla 3D Time-of-Flight~(ToF) Magnetic Resonance Angiography (MRA) images. Then it is used to perform a five-fold cross-validation. The generalisation performance of the method is evaluated using the other unseen volumes in the dataset. It is observed that the proposed method with multi-axes MIP loss produces better quality segmentations with a median Dice of $80.245 \pm 0.129$. Also, the method with single-axis MIP loss produces segmentations with a median Dice of $79.749 \pm 0.109$. Furthermore, a visual comparison of the ROIs in the predicted segmentation reveals a significant improvement in the continuity of the vessels when MIP loss is incorporated into training.  
\end{abstract}


\end{frontmatter}


\section{Introduction}

Segmentation of vessels on 7T magnetic resonance angiography (MRA) is one of the most important tasks in the analysis of biomedical images, as it provides essential information for the diagnosis and treatment of cerebrovascular diseases. An enhanced signal-to-noise ratio offered by 7T MRA allows for superior visualisation of cerebral vessels, revealing a higher proportion of small vessels compared to 1.5T or 3T MRAs \cite{avadiappan2020fully}. The intricacies of small vessel structures complicate the segmentation process. Manual segmentation is labour-intensive and often leads to errors due to the challenging nature of the task. Machine learning (ML), specifically deep learning (DL), has shown promise in automating and improving the accuracy of vessel segmentation \cite{avadiappan2020fully}. Despite its potential, deep learning in 7T MRA vessel segmentation is hindered by the need for extensive and expert-driven annotations. The manual segmentation process, necessary for training these models, is time-consuming and labour-intensive, especially given the intricacy of vessel structures in 7T MRAs. DS6 \cite{Chatterjee_2022}, a semi-supervised deep learning approach, attempts to mitigate this issue by learning from a small dataset with noisy annotations. Although this method successfully segments vessels as small as one to two voxels in diameter, it does not always produce segmentations that preserve the continuity of the vessels. Current research focuses on improving the continuity of vessels by considering the spatial correlation of pixels across dimensions.

For high-resolution 3D 7T MRA, creating ground truth annotations without imperfections is challenging. Usually, manual, semi-automatic, or classic vesselness segmentations are used to create training labels. However, these imperfections in training labels can reduce the performance of the trained deep learning network. In particular, if single voxels or small clusters are missed in the training labels, discontinuities in the deep learning-based vessel segmentation can occur. To overcome these challenges, this research attempts to take advantage of the inherent property of maximum intensity projection~(MIP) to compress the 3D training annotations into 2D projections. Conventional maximum intensity projections~(MIPs) are often used to quickly assess the vasculature in a single 2D image (or projection), instead of browsing through the entire 3D dataset. Although this reduces the dimensionality of the data, it does compensate for small clusters of missing labels. Using the MIP of the training label as an additional source to create a loss term, the authors expect an improvement in vessel continuity in the resulting DL segmentations.

\subsection{Related Work}

\subsubsection{Vessel segmentation using manual and non-DL based methods}
To distinguish between vessels and non-vessels, experts assign each voxel a value of 0 for non-vessels and 1 for vessels. The detection of the lenticulostriate arteries~(LSA) is one of the key tasks in this study, which includes the annotation of large and small vessels. Unlike detecting large vessels, perceiving the gaps between these vessels consisting of extremely small vessels~(LSA) is difficult for the human eye. Therefore, manual segmentation procedures are biased towards the perspectives and expertise of the individual performing the annotations. The need to annotate a large number of voxels precisely to render segmentations of the 3D volume makes the task time-consuming.

The Frangi~\cite{Frangi1998} filter is designed to enhance blood vessels and other tubular structures with the eventual goal of vessel segmentation by improving contrast and reducing noise. The approach is based on Hessian eigenvalues, which are instrumental in the vessel contrast enhancement and suppression of non-vascular structures. However, this method requires significant parameter tuning to identify small vessels of interest. Occasionally, these parameters need to be manually fine-tuned according to each dataset and volume.

The 'Openly available sMall vEsseL sEgmenTaTion pipelinE'~(OMELETTE)\footnote{The Openly available sMall vEsseL sEgmenTaTion pipelinE~(OMELETTE): \href{https://gitlab.com/hmattern/omelette}{https://gitlab.com/hmattern/omelette}}~\cite{OMELETTE} method focuses on segmenting images based on thresholds. The voxels above a certain threshold are considered vessels, and the rest are considered as background. Hysteresis thresholding is employed as it helps maintain vessel continuity by considering voxels above the lower threshold if they are connected to the vessels with higher thresholds. Additionally, Jerman's filter is used to apply Jerman's vessel response function, which is based on the volume ratio of the Hessian matrix eigenvalues.

\subsubsection{Vessel segmentation using deep learning techniques}
Convolutional Neural Networks~(CNN) have been extensively used for computer vision and image processing tasks. The high-level feature representations learnt using such networks can be efficiently used as segmentation boundaries. However, the biomedical image segmentation task presents the challenge of learning such representations using limited weak annotations. UNet~\cite{Ronneberger2015UNet} architecture proposes an end-to-end trainable network with a contracting path to learn high-resolution context information, followed by a symmetric expanding path that produces more precisely localised segmentation. 

UNet-based architectures have been proven to be efficient in the task of segmenting vessels. One such network is the UNet with Multi-Scale Deep Supervision~(UNet MSS)~\cite{Zeng2017UNetMSS,zhao2019multi}. Zeng et.al. proposed a multi-scale loss to learn discriminative features at every level and computed the overall loss as the sum of losses at each up-sampling scale of the expansion path of the UNet. Using this architecture as the backbone, Chatterjee et.al.~\cite{Chatterjee_2022} proposed a semi-supervised deformation-aware learning approach for vessel segmentation with noisy labels. The limited annotated samples were augmented by subjecting them to random elastic deformations. The deformed samples were trained using a Siamese architecture based on UNet~\cite{Ronneberger2015UNet} and UNet-MSS~\cite{Zeng2017UNetMSS,zhao2019multi} models. The approach was based on the hypothesis that learning features at different scales help segment vessels of different sizes and that deformation awareness improves consistency given a small set of noisy samples. 

\subsubsection{Maximum intensity projection}
Maximum intensity projection~(MIP) is used to visualise hyperintense structures in a 3D volume as a 2D projection, where, for each projection trace, only the voxel with the highest intensity is shown in the final 2D MIP. \citep{GAN2004195} hypothesises that a higher proportion of vessel structure is apparent in the MIPs as opposed to the 3D volumes and this can be exploited in cerebrovascular segmentation. MIPs have also been instrumental in the detection of pathologies. \citep{Adachi2020MIP} and \citep{Hu2021MIP} demonstrated the use of MIPs of dynamic contrast-enhanced MRIs in detecting and classifying breast lesions. Furthermore, studies by \citep{ZHENG2020105620} and \citep{CAO2020696} have shown that MIPs can be instrumental in detecting pulmonary nodules and qualitative analysis of intracranial vascularity. In the current study, the authors hypothesise that the MIP of the 3D MRA annotations can be used to improve the UNet-MSS~\cite{zhao2019multi,Chatterjee_2022} network's perception of vessel continuity.

\subsection{Contributions}
This attempts to tackle the problem of vessel continuity in deep learning-based segmentation models by introducing a novel approach by incorporating maximum intensity projection (MIP) as an additional loss criterion. Two versions of the proposed loss term have been explored here and have been employed on two different deep learning models and evaluated for overall segmentation quality, underlying vasculature, and vessel continuity. This advancement has significant potential to improve the precision and reliability of vessel segmentation in neuroimaging, thereby contributing to the better diagnosis and treatment of cerebrovascular diseases, especially small vessel disorders.

\section{Methodology}

\begin{figure*}
\centering
\includegraphics[scale=0.44]{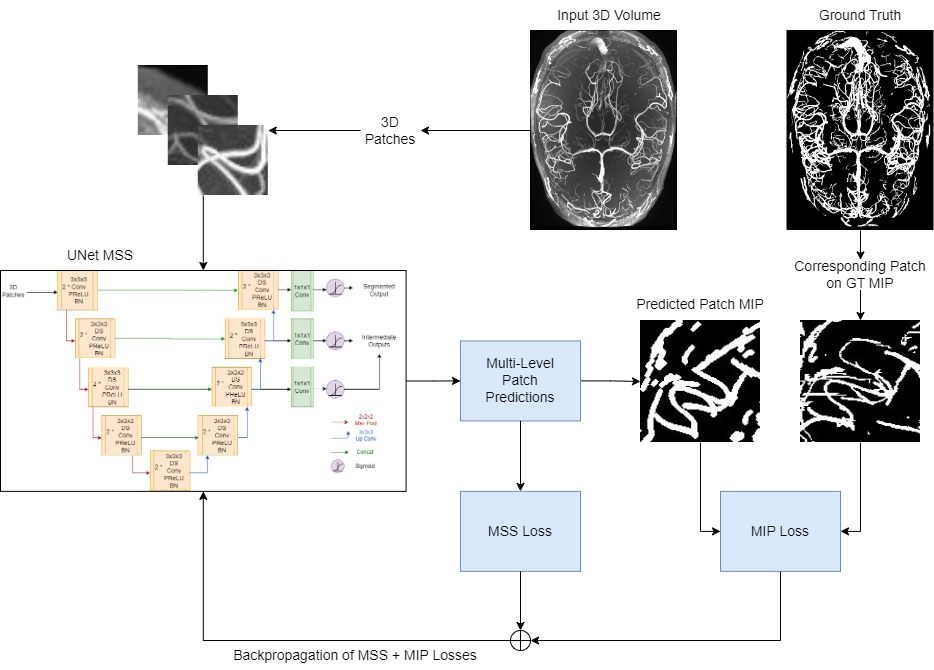}
\caption{The Modified UNet MSS network is trained on 3D patches created from the input volume. MIPs of multi-level predictions are then compared against their corresponding patches on the MIP of the ground truth to compute MIP loss. The MSS loss is computed by comparing multi-level patch predictions with the corresponding label patches. A weighted sum of the two losses is then backpropagated.}
\label{fig:Proposed Training Pipeline}
\end{figure*}

\subsection{Proposed Approach: SPOCKMIP}

This paper proposes SPOCKMIP\footnote{SPOCKMIP: \textbf{S}egmentation \textbf{P}recision \textbf{O}ptimised with \textbf{C}ontinuity \textbf{K}nowledge using \textbf{M}aximum \textbf{I}ntensity \textbf{P}rojection} method, that uses the same architecture of UNet-MSS model from the DS6 research~\cite{Chatterjee_2022} with a replacement of the activation function from ReLU
to LeakyReLU, and enhances the patch-based training pipeline by introducing the MIP comparisons as an additional loss term, as shown in the Fig. \ref{fig:Proposed Training Pipeline}. 
The MIPs of the predictions for each patch at each level of UNet-MSS are computed. The predicted patch MIPs are then compared with their corresponding patches in the respective label MIPs to evaluate the MIP loss, $L_{MIP}(\theta)$ as shown in Eq~(\ref{eq:4}) and the Fig. \ref{fig:MIP Comparisons}.

\subsubsection{Maximum intensity projection loss along the slice-dimension}
In addition to the Multi-Scale Supervision~(MSS) Loss, the spatial continuity of the vessels along the z-axis (i.e. the slice dimension) is incorporated into the learning in the form of the MIP loss. Eq.~\ref{eq:1} represents the total loss, which is a weighted sum of the MSS loss $L_{MSS}(\theta)$ and the MIP loss $L_{MIP}(\theta)$ with weight parameter $\mu$ and network parameter $\theta$. Eq.~\ref{eq:2} represents the MSS loss where $m$ refers to the total up-sampling scales, and $\alpha_{_i}$ is the weight assigned to the loss at a specific up-sampling level. Eq.~\ref{eq:3} represents the MIP loss that is calculated by comparing the MIP of the predicted segmentation of the patch $\hat{y}$ against the subset of the MIP of the label segmentation $Y$ encompassing the patch. The Focal Tversky loss~\cite{abraham2018novel} is used as the loss function for calculating all losses. 
\begin{equation} \label{eq:1}
 Loss(\theta) = \mu L_{MSS}(\theta) + (1 - \mu) L_{MIP}(\theta) 
\end{equation}
\begin{equation} \label{eq:2}
L_{MSS}(\theta) = {1 \over m} { \sum_{i=1}^{m} \alpha {_i} l_{i} (\theta)}
\end{equation}
\begin{equation} \label{eq:3}
L_{MIP}(\theta) = {1 \over m} { \sum_{i=1}^{m} \alpha {_i} lmip_{i} (\theta)}
\end{equation}
\begin{equation} \label{eq:4}
 lmip(\theta) = loss(MIP(\hat{y}),y \subseteq MIP(Y))
\end{equation}

\subsubsection{Cumulative maximum intensity projection loss across multiple axes} 
\begin{figure*}
     \centering
     \begin{subfigure}[b]{0.40\textwidth}
         \centering
         \includegraphics[width=\textwidth]{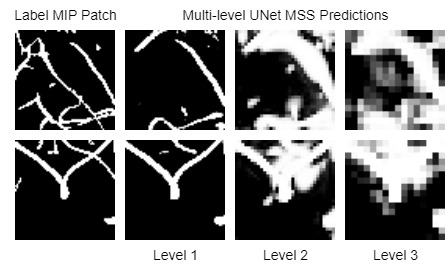}
         \caption{}
         \label{fig:MIP Comparisons}
     \end{subfigure}
     \hfill
     \begin{subfigure}[b]{0.40\textwidth}
         \centering
         \includegraphics[width=\textwidth]{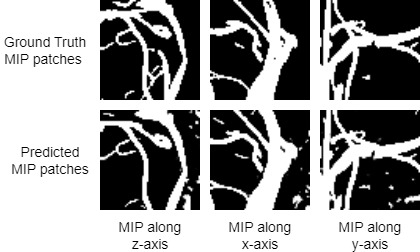}
         \caption{}
         \label{fig:Multi-axis Mip Comparisons}
     \end{subfigure}
        \caption{(a) MIP loss is calculated by comparing the MIP of the network's predictions at each level for a patch with the corresponding patch on the MIP of the ground truth segmentation. (b)  Overall MIP loss is calculated by comparing the MIPs along three perceivable axes of the network's predictions at each level for a patch with the corresponding patches on the respective MIPs along multiple axes of the ground truth segmentation.}
        \label{fig:MIPComaprisons}
\end{figure*}

The authors propose an additional hypothesis that the continuity of vessel structures can be better perceived by analysing MIPs of the volume across multiple axes. This is achieved by comparing the MIP of the network's 3D patch predictions against the corresponding patches of MIPs of 3D labels across three different views, as shown in Fig.~\ref{fig:Multi-axis Mip Comparisons}. Therefore, the overall MIP loss is calculated as an equally weighted sum of the MIP loss along each axis, as shown in Eq.~\ref{eq:5}, where $\beta$ represents the weight coefficient of the MIP loss across the $x$, $y$ and $z$ axes.
\begin{equation} \label{eq:5}
L_{MIP}(\theta) = {1 \over m} \beta { \sum_{i=1}^{m} \alpha {_i} (lmip_{x,i} (\theta)+lmip_{y,i} (\theta)+ lmip_{z,i} (\theta))}
\end{equation}

\subsubsection{Hypothesis}
The authors hypothesise that the proposed modifications, which take into account the MIP of the volume, enhance the segmentation of small vessels and improve vessel continuity. This hypothesis is evaluated and the performance of the proposed methods is compared against the baseline approaches including UNet~\cite{Ronneberger2015UNet} and UNet-MSS~\cite{Zeng2017UNetMSS} both in terms of segmentation quality using ROI comparisons and a quantitative evaluation with a set of standard vessel segmentation metrics.  

\subsection{Datasets and Labels}

\begin{figure}
\centering
\includegraphics[width=0.2\textwidth]{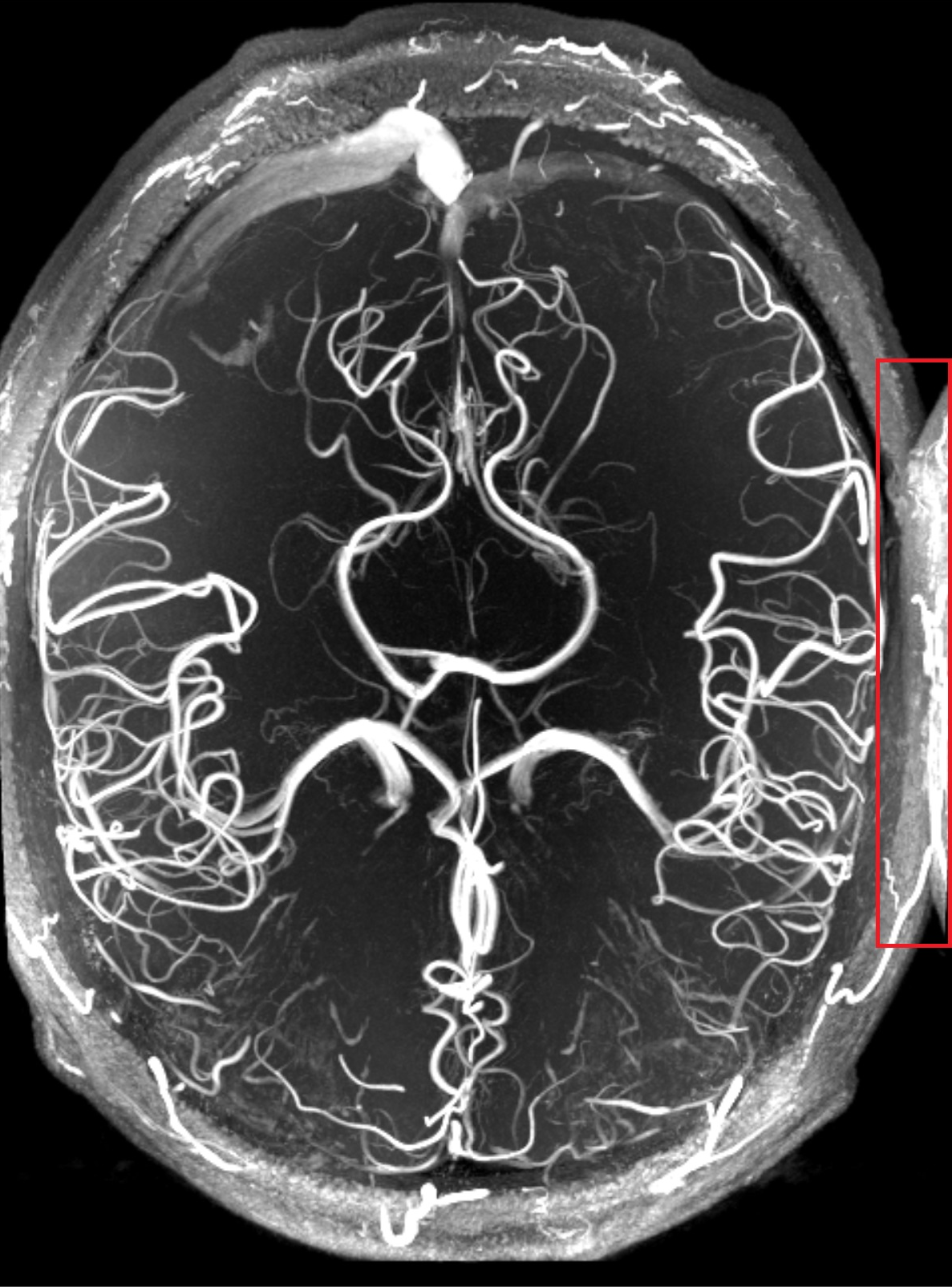}
\caption{Wrap-around artefact}
\label{fig:wrap-around}
\end{figure}

The proposed methodology was evaluated using the \textit{StudyForrest}\footnote{\href{https://studyforrest.org}{Source: StudyForrest.org}} dataset~\cite{SF7-TeslaDataset}, comprising 7T MRA volumes of the brain acquired using 3D multi-slab Time-Of-Flight~(TOF) Magnetic Resonance Angiography~(MRA) with a resolution of 300$\mu$m of 20 participants. These volumes were divided into three subsets: the training set included 12 volumes, the validation set included four volumes, and the testing set included four volumes. 
The \textit{StudyForrest} dataset includes two volumes with phase wrap-around artefacts~(Fig. \ref{fig:wrap-around}), which are regularly observed MR artefacts that occur when the dimensions of the body part being imaged exceed the defined Field of View~(FOV). Those two volumes with these artefacts were discarded from the training as they hindered learning by increasing noise, and they were used for additional evaluations.

\subsubsection{Label preparation}
The labels for the test volumes were manually annotated and verified by a neurologist, while the labels for the training and validation sets were created in a semi-automated fashion using Ilastik~\cite{Berg2019} and the 3D slicer~\cite{Fedorov20123DSlicer}. After annotating the volumes using Ilastik, the MIP of the original volume along the slice dimension was used to validate the accuracy. The resulting segmentation had thicker labels along with abundant noise in the skull region. The label thickness was reduced by annotating the outlining pixels of the vessel labels by non-vessel labels. To minimise skull noise, the noisy skull region was removed by retaining the prominent vessels using 3D slicer. Skull vessels were annotated separately in Ilastik and combined with the prominent vessels mentioned earlier. These methods resulted in accurate segmented volumes with minimal noise. The area opening and area closing morphological methods provided by the \textit{scikit-image} library were then applied to reduce noise and improve vessel continuity, respectively. The area opening operation was fine-tuned with an area threshold of seven and a connectivity value of two, followed by an area closing operation with an area threshold of sixty and a connectivity value of four.

\subsection{Experimental Setup}
A 5-fold cross-validation is performed on 14 volumes, where each fold comprises three volumes for validation, while the rest are used for training. The generalisation performance of the thus trained models is evaluated using the held-out test set comprising the remaining four unseen volumes (three volumes were free of wrap-around artefacts, while one volume had such artefact). 3D MRA volumes with dimensions of 480x640x163 were converted to 3D patches of 64\textsuperscript{3}. 8000 such patches from the 11 training volumes are randomly selected on each epoch for training. All experiments were performed with a 32GB Nvidia Tesla V100-SXM2 GPU with 10 CPUs and 60 GB RAM. 

\begin{figure*}[ht]
\centering
\includegraphics[width=0.6\textwidth]{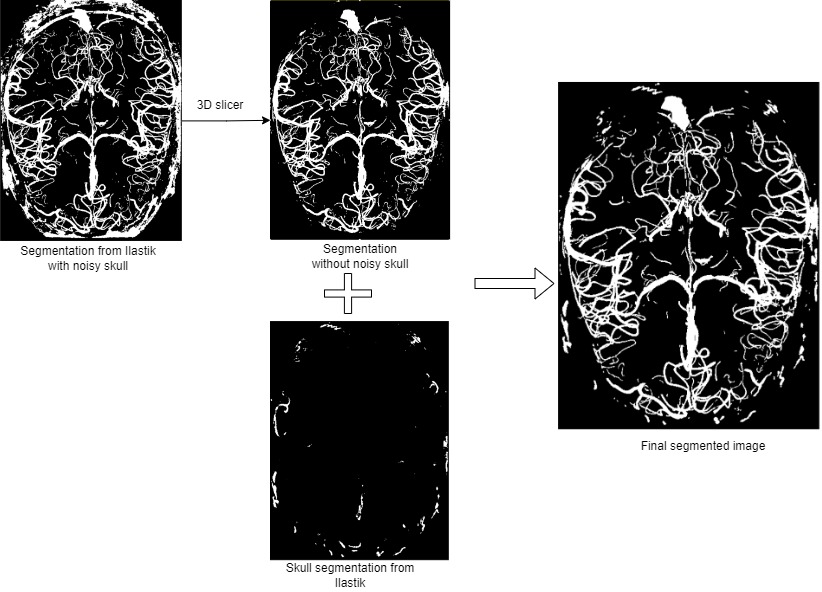} 
\caption{Label creation using Ilastik and 3D slicer}
\label{fig:Label Creation}
\end{figure*}

SRDataset\footnote{SRDataset is a PyTorch dataset extension used to create patches from a 3D volume and lazy load them using a TQDM data loader. The code is referred from \href{https://github.com/soumickmj/FTSuperResDynMRI/blob/main/utils/datasets.py}{https://github.com/soumickmj/FTSuperResDynMRI}} is used to prepare the dataset of patches for training and validation. The dataset was adapted to identify, for each patch, the corresponding location of the patch on the MIP of the segmentation label using the patch coordinates. Thus, it returns the patch, the corresponding label patch, and the corresponding label MIP patch on each load. All experiments were performed with a learning rate of 0.0001 over 50 epochs. Focal Tversky Loss~\cite{abraham2018novel} was used as a loss function for the calculation of both supervised loss~(MSS loss) and MIP loss. These were optimised during training using the Adam optimiser~\cite{kingma2017adam}. 
Additionally, Automatic Mixed Precision~(AMP)\footnote{Using \href{https://github.com/NVIDIA/apex}{Nvidia Apex}} and gradient clipping techniques were employed to reduce memory requirements and to prevent exploding gradients. 

\subsubsection{Hyperparameters}

\begin{table}
\centering
\resizebox{0.45\textwidth}{!}{%
\begin{tabular}{lr}
\hline
\textbf{Hyperparameter} & \textbf{Value} \\
\hline
\textbf{Batch Size} & 15 \\ 
\textbf{Patch Size} & 64 \\ 
\textbf{Number of Epochs} & 50 \\ 
\textbf{Learning Rate} & 0.0001 \\ 
\textbf{Stride Dimensions} & 16x32x32 (Depth x Width x Length) \\ 
\textbf{Samples Per Epoch} & 8000 \\ 
\textbf{MSS loss Coefficient ($\mu$)} & 0.7 \\ 
\textbf{MIP loss Coefficient ($1 - \mu$)} & 0.3 \\ 
\textbf{MSS level Coefficient ($l_{i}$)} & [1, 0.66, 0.34] on levels 1, 2, and 3 respectively of UNet MSS~\cite{Zeng2017UNetMSS} \\ 
\textbf{MIP axis Coefficient ($\alpha_{i}$)} & 0.33 (equal weightage) \\ 
\hline
\end{tabular}%
}
\caption{The hyperparameters and their optimal values are based on initial experiments.}
\label{tab:HyperParams}
\end{table}

Table~\ref{tab:HyperParams} shows the set of different hyperparameters and their optimal values selected based on initial experiments, memory constraints, and the previous study~\cite{Chatterjee_2022}. The large 3D MRA volumes were divided into 3D patches of equal dimensions to facilitate memory-efficient learning while also increasing the number of samples per epoch. In each training iteration, a set of patches was chosen from different training volumes at random and supplied to the network. Taking into account the underlying limitation of the GPU and the size of the dataset used for the experiments, the batch size was set to 15, and the patch dimensionality was set to $64^3$. 

3D patches were created using SRDataset, where the coordinates of the patches are determined by traversing the 3D volume with an arbitrary stride. The dimensions of this stride were set to 16x32x32~(depth x width x length) to allow enough overlap. The patch coordinates returned by the tool were used to lazy load the set of patches during training, and each training volume was divided into more than 10,000 overlapping patches. To mitigate overfitting, 8,000 randomly selected patches are used for each training epoch.

The proposed overall loss is calculated as a weighted sum of multi-scale supervision loss~(MSS loss) and MIP loss. Optimising the network weights by minimising this multi-objective loss often results in exploding gradients. To allow the model to explore the local solution landscape of each loss term, a lower learning rate of 0.0001 was selected compared to the previous study~\cite{Chatterjee_2022}. It was observed that the convergence of the single- and multi-axes experiments occurred after 42 epochs. Therefore, the number of epochs is set to 50.

\subsubsection{Loss coefficients}
The complexity of learning the optimal objective function using deep learning models increases with the number of optimisation criteria. Here, the network is essentially optimising two objectives: voxel similarity loss~(MSS loss) and maximum intensity projection loss~(MIP loss). The proposed approach for this multi-objective optimisation is to optimise the weighted sum of the two losses. The hyperparameter $\mu$ in Eq.~(\ref{eq:1}) is added as an additional network parameter with an initial value of 0.7, which is selected based on experimental results. It was found that the optimal value of the parameter learnt by the network was found to be 0.68. Further experiments showed that the learning suffered from under-fitting with reduced importance on MSS loss i.e., with $\mu < 0.7$. On the other hand, a substantial increase in over-segmentation was observed with increasing weight on MIP loss, i.e., $\mu > 0.7$. The weight coefficient $\alpha$ associated with the MIP loss of each axis \textit{i} in Eq.~(\ref{eq:5}) was set to 0.33, implying equal importance for each axis. Several experiments were performed by assigning different importance to MIP loss at each axis, including a random assignment. These resulted in over-segmentation, ranging from overlapping vessel boundaries to identifying everything as vessels. The loss coefficient $\beta$ in Eq. \ref{eq:5} was set to 0.3, similar to $1 - \mu$ in Eq. \ref{eq:1}. 

\paragraph{Single-axis MIP experiments} The coefficient $\mu$ was set to 0.7 after a set of initial experiments. In addition, the hyperparameter was added as a parameter to the network, and the optimal value was found to be 0.68. ReLU activation was replaced by LeakyReLU to improve the optimisation of the combined losses. A higher weight was assigned to the MSS loss to learn the underlying vessel features.

\paragraph{Multi-axis MIP Experiments} The SRDataset\textsuperscript{2} was adapted to compute the MIP of the labels across three dimensions of perception. The corresponding patches were identified and returned on each data load, along with the patch and its corresponding label patch. The MIP loss was computed as the sum of MIP loss along each axis, as shown in Eq.~(\ref{eq:5}). Equal weights were assigned to the individual axis losses, and the coefficient $\beta$ was set to 0.33 after initial experiments. The weight coefficient $\mu$ was set to 0.7 for the overall sum of losses, as in Eq.~(\ref{eq:1}). 

\subsection{Evaluation}
The segmentation results are quantitatively evaluated against the manually segmented ground truth in terms of the overall segmentation score using Dice Coefficient~(Dice), Area under ROC Curve~(AUC), and Sensitivity. In addition, the underlying structure of the vessels and the parts of the vessels identified are evaluated using Volumetric Similarity Coefficient~(VS), Mutual Information~(MI), and Mahalanobis Distance~(MHD). The Dice Coefficient and Volumetric Similarity are instrumental in evaluating the overlap and volumetric precision of segmentations, while the AUC and Sensitivity provide critical insights into the model's discriminative power and its efficacy in accurately identifying true positives. Furthermore, the employment of Mutual Information and Mahalanobis Distance metrics captures both structural similarity and geometric accuracy. This judicious selection of diverse metrics ensures a robust and comprehensive assessment of segmentation quality, thereby significantly enhancing the model's reliability and applicability in the field. 

The Dice Coefficient is a measure of overlap between two sets, \(A\) and \(B\). For segmentation tasks, it can be defined as:
\begin{equation} \label{eq:dice} 
\text{Dice}(A, B) = \frac{2|A \cap B|}{|A| + |B|} 
\end{equation}
where \(A\) and \(B\) are the sets of predicted and ground-truth segments, and \(|*|\) is the number of elements in that particular set. The Area Under the ROC Curve is a performance measurement for classification tasks (segmentation tasks can be considered as a pixel-wise classification task) at various threshold settings. The ROC is a probability curve and the AUC represents the degree or measure of separability. It is computed as:

\begin{equation} \label{eq:auc} 
\text{AUC} = \int_{0}^{1} \text{TPR}(t) \, d\text{FPR}(t) 
\end{equation}
where \(\text{TPR}(t)\) and \(\text{FPR}(t)\) are the true positive and false positive rates at threshold \(t\). Sensitivity, or recall, is the proportion of true positives correctly identified by the model and is computed as:

\begin{equation} \label{eq:sensitivity}  
\text{Sensitivity} = \frac{\text{TP}}{\text{TP} + \text{FN}} 
\end{equation}
where \(\text{TP}\) and \(\text{FN}\) are the number of true and false positives. The Volumetric Similarity Coefficient measures the similarity in volume between the predicted segmentation and the ground truth using:

\begin{equation} \label{eq:volsim} 
\text{VS} = 1 - \frac{|V_A - V_B|}{V_A + V_B} 
\end{equation}
where \(V_A\) and \(V_B\) are the volumes of the predicted and ground-truth segments. Mutual Information (MI) measures the amount of information obtained about one random variable through the other random variable. For segmentation, it can be expressed as:

\begin{equation} \label{eq:mi} 
\text{MI}(A, B) = \sum_{a \in A} \sum_{b \in B} p(a, b) \log \frac{p(a, b)}{p(a) p(b)} 
\end{equation}
where \(p(a, b)\) is the joint probability distribution of \(A\) and \(B\), while \(p(a)\) and \(p(b)\) are the marginal probability distributions. The Mahalanobis Distance is a measure of the distance between a point and a distribution. For segmentation, it is used to measure the distance between the predicted segmentation and the ground truth:

\begin{equation} \label{eq:mhd}  
\text{MHD}(A, B) = \sqrt{(A - B)^T \Sigma^{-1} (A - B)} 
\end{equation}
where \(A\) and \(B\) are the vectors of predicted and ground-truth segment coordinates, and \(\Sigma\) is the covariance matrix of the distribution.

To facilitate the calculation of the above-mentioned metrics, EvaluateSegmentation\footnote{\href{https://github.com/Visceral-Project/EvaluateSegmentation}{EvaluateSegmentation: A comprehensive tool to compare two medical volumes.}}~\cite{EvaluateSeg} tool was employed.

\section{Results}

\subsection{Quantitative Evaluation}
A test set, comprising four 7T MRA volumes, is used to evaluate the 5-fold cross-validated models of each approach. One of the volumes in this set contains wrap-around artefacts, which affect the objective quantitative evaluation of the segmentation. Therefore, the resulting comparisons are presented in two categories: with and without the volume containing the aforementioned artefacts.

\subsubsection{On test set without wrap-around artefacts}
\begin{figure}
     \centering
     \begin{subfigure}[b]{0.47\textwidth}
         \centering
         \includegraphics[width=\textwidth]{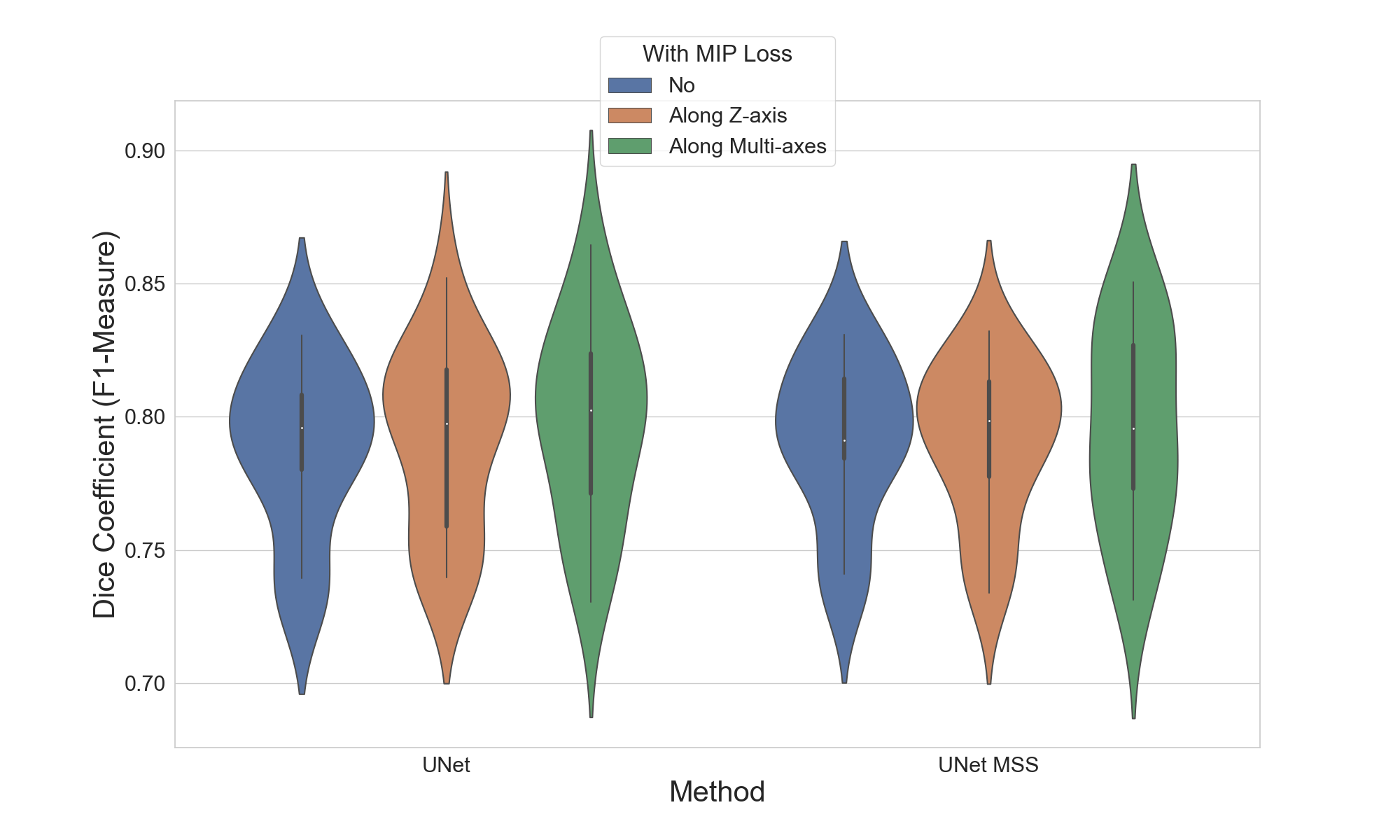}
         \caption{Comparison of overall segmentation scores of proposed methods evaluated using the Dice Coefficient metric on the test set without wrap-around artefacts.}
         \label{fig:Dice Violin without WAA}
     \end{subfigure}
     \hfill
     \begin{subfigure}[b]{0.47\textwidth}
         \centering
         \includegraphics[width=\textwidth]{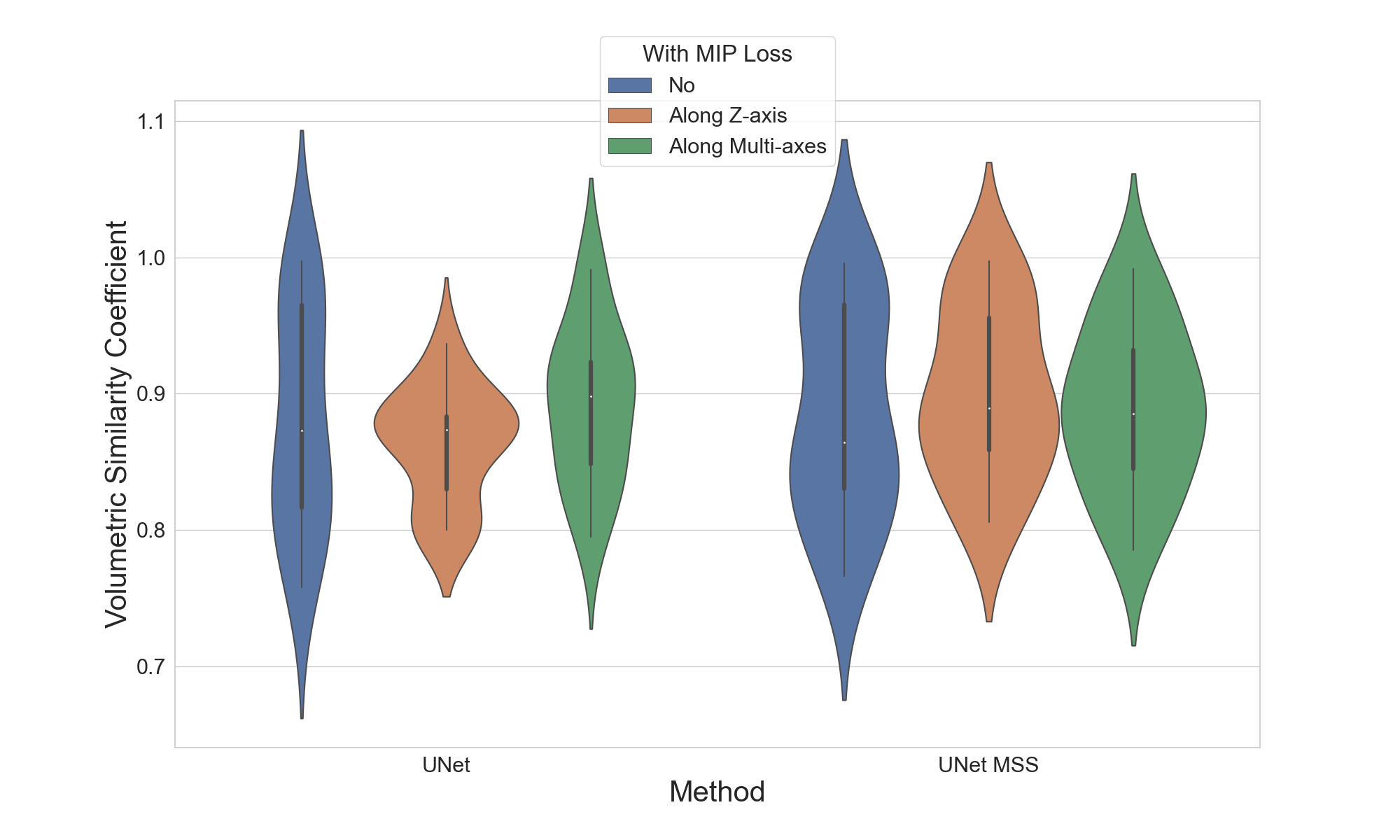}
         \caption{Quantitative comparison of underlying vasculature identified by the proposed methods, evaluated using the Volumetric Similarity Coefficient metric on the test set without wrap-around artefacts.}
         \label{fig:VS Violin without WAA}
     \end{subfigure}
        \caption{Violin plots comparing the performance of proposed methods and baselines over three test volumes, excluding the volume with wrap-around artefacts, across five folds.}
        \label{fig:Violin Plot without WAA}
\end{figure}

The median and variance over 15 segmentation results obtained from a 5-fold cross-validation (three volumes of the test set without wrap-around artefacts, evaluated over five folds, resulting in 15 segmentation results in total) are used to compare the performance of the methods. These results are reported in Table \ref{tab:Metrics Comparison without WAA} and presented using violin plots in Fig. \ref{fig:Violin Plot without WAA}.

\begin{table}
  \centering
    \begin{subtable}[t]{0.45\textwidth}
      \centering
      \resizebox{\textwidth}{!}{%
        \begin{tabular}{@{}cccc@{}}
\toprule
\textbf{Methods / Metrics} & \textbf{Dice Coefficient} & \textbf{Area under ROC Curve} & \textbf{Sensitivity} \\ \midrule
UNet & $79.585 \pm 0.092$ & $0.84869 \pm 0.00160$ & $0.69762 \pm 0.00648$ \\
UNet MSS & $79.130 \pm 0.085$ & $0.86042 \pm 0.00140$ & $0.72123 \pm 0.00569$ \\
\textbf{UNet MIP} & $79.749 \pm 0.109$ & $0.85595 \pm 0.00070$ & $0.71262 \pm 0.00279$ \\
\textbf{UNet MSS MIP} & $79.857 \pm 0.080$ & $0.86447 \pm 0.00100$ & $0.72939 \pm 0.00403$ \\
\textbf{UNet mMIP} & $\textbf{80.245} \pmb{\pm} \textbf{0.129}$ & $\textbf{0.86672} \pmb{\pm} \textbf{0.00127}$ & $\textbf{0.73403} \pmb{\pm} \textbf{0.00512}$ \\
\textbf{UNet MSS mMIP} & $79.577 \pm 0.136$ & $0.86134 \pm 0.00118$ & $0.72413 \pm 0.00474$ \\ \bottomrule
\end{tabular}%
        }
         \caption{Overall Segmentation Scores}
        \label{tab:OverallScoreWWAA}
    \end{subtable}
    
    \vspace*{5mm}
    \centering
    
    \begin{subtable}[t]{0.45\textwidth}
      \centering
      \resizebox{\textwidth}{!}{%
        \begin{tabular}{@{}cccc@{}}
\toprule
\textbf{Methods / Metrics} & \textbf{Volumetric Similarity} & \textbf{Mutual Information} & \textbf{MHD} \\ \midrule
UNet & $0.87312 \pm 0.00634$ & $0.04758 \pm 0.00002$ & $0.06812 \pm 0.00051$ \\
UNet MSS & $0.86421 \pm 0.00568$ & $0.04783 \pm 0.00002$ & $0.07158 \pm 0.00082$ \\
\textbf{UNet MIP} & $0.87358 \pm 0.00163$ & $0.04790 \pm 0.00001$ & $\textbf{0.06053} \pmb{\pm} \textbf{0.00037}$ \\
\textbf{UNet MSS MIP} & $0.88928 \pm 0.00364$ & $0.04867 \pm 0.00001$ & $0.06625 \pm 0.00059$ \\
\textbf{UNet mMIP} & $\textbf{0.89829} \pmb{\pm} \textbf{0.00312}$ & $0.04900 \pm 0.00002$ & $0.06629 \pm 0.00033$ \\
\textbf{UNet MSS mMIP} & $0.88557 \pm 0.00338$ & $\textbf{0.04944} \pmb{\pm} \textbf{0.00001}$ & $0.06159 \pm 0.00076$ \\ \bottomrule
\end{tabular}%
        }
         \caption{Quantitative evaluation of underlying vasculature}
        \label{tab:VasculatureWWAA}
    \end{subtable}
\caption{Metrics comparisons of 15 segmentation results of test volumes, excluding the volume with wrap-around artefacts, over 5-fold cross-validation.}
\label{tab:Metrics Comparison without WAA}
\end{table}

The overall segmentation scores, presented in Table~\ref{tab:OverallScoreWWAA}, show improvements with the incorporation of MIP loss compared to their baseline counterparts. It is evident from the Dice score comparisons that the UNet mMIP method clearly outperforms the baselines, with a median Dice score of $80.245 \pm 0.129$. The AUC and sensitivity comparisons also show that the proposed MIP loss improves segmentation performance. The improvements are considerably greater in the case of the baseline UNet compared to the UNet MSS. The multi-axes UNet mMIP method outperforms its baseline, with a median AUC of $0.867 \pm 0.001$.

Furthermore, the volumetric comparisons of the identified vasculature are presented in Table~\ref{tab:VasculatureWWAA}, and they also demonstrate improvements over the baselines with the addition of MIP loss. The UNet model with multi-axes MIP loss, UNet mMIP, outperforms the baselines and other proposed MIP-based methods, with a median score of $0.898 \pm 0.003$. The comparison of the MHD metric shows that the UNet MSS mMIP method exploits voxel correlations better than the baselines and other proposed methods. 

The study proposes two methods for the incorporation of vessel structure information available in the MIP. This includes single-axis MIP-based methods~(UNet MIP and UNet MSS MIP) and multi-axes MIP-based methods~(UNet mMIP and UNet MSS mMIP). The authors hypothesised that projecting the 3D volume along all three perceivable axes allows the network to learn better spatial correlation of the voxels compared to single-axis projection. Although this hypothesis is validated in the case of the baseline UNet, a reduction is observed in the overall segmentation scores of the UNet MSS with the addition of MIP information from multiple axes. The qualitative evaluation discussed in a later section explains this observation as a consequence of increased over-segmentation. The quantitative evaluation of the underlying vasculature shown in Table~\ref{tab:VasculatureWWAA} reveals a considerable improvement in the UNet segmentation results with multi-axis MIP loss as opposed to single-axis MIP loss evident in both VS and MI scores. The performance of the UNet MSS remains marginally similar with the incorporation of single- and multi-axes MIP loss.

Overall, it can be concluded that MIP loss (single- or multi-axes) improves the segmentation performance of both models on all six metrics. However, the conclusion regarding the improvement using multi-axes MIP over single-axis MIP depends on the model. The authors hypothesise that as the UNet MSS already has additional regularisation in terms of the additional loss terms, adding multi-axes MIP might be over-regularising the model - ending up losing performance compared to the single-axis MIP loss.

\subsubsection{On test volume with wrap-around artefacts}

\begin{table}
  \centering
    \begin{subtable}[t]{0.45\textwidth}
      \centering
      \resizebox{\textwidth}{!}{%
        \begin{tabular}{@{}cccc@{}}
\toprule
\textbf{Methods / Metrics} & \textbf{Dice Coefficient} & \textbf{Area under ROC Curve} & \textbf{Sensitivity} \\ \midrule
UNet & $63.080 \pm 0.109$ & $0.73868 \pm 0.00122$ & $0.47775 \pm 0.00499$ \\
UNet MSS & $64.070 \pm 0.048$ & $0.74810 \pm 0.00070$ & $0.49677 \pm 0.00285$ \\
\textbf{UNet MIP} & $64.177 \pm 0.037$ & $0.76815 \pm 0.00022$ & $0.53774 \pm 0.00088$ \\
\textbf{UNet MSS MIP} & $64.821 \pm 0.040$ & $0.76566 \pm 0.00037$ & $0.53248 \pm 0.00152$ \\
\textbf{UNet mMIP} & $64.357 \pm 0.125$ & $0.75854 \pm 0.00138$ & $0.51805 \pm 0.00563$ \\
\textbf{UNet MSS mMIP} & $\textbf{65.946} \pmb{\pm} \textbf{0.084}$ & $\textbf{0.77267} \pmb{\pm} \textbf{0.00067}$ & $\textbf{0.54650} \pmb{\pm} \textbf{0.00271}$ \\ \bottomrule
\end{tabular}%
        }
         \caption{Overall Segmentation Scores}
        \label{tab:OverallScoreWAA}
    \end{subtable}
    
    \vspace*{5mm}
    \centering
    
    \begin{subtable}[t]{0.45\textwidth}
      \centering
      \resizebox{\textwidth}{!}{%
        \begin{tabular}{@{}cccc@{}}
\toprule
\textbf{Methods / Metrics} & \textbf{Volumetric Similarity} & \textbf{Mutual Information} & \textbf{MHD} \\ \midrule
UNet & $0.70075 \pm 0.01043$ & $0.03268 \pm 0.00001$ & $0.29559 \pm 0.00142$ \\
UNet MSS & $0.72650 \pm 0.00660$ & $0.03352 \pm 0.00001$ & $0.25334 \pm 0.00184$ \\
\textbf{UNet MIP} & $0.76243 \pm 0.00171$ & $0.03429 \pm 0.00000$ & $0.25282 \pm 0.00079$ \\
\textbf{UNet MSS MIP} & $0.78264 \pm 0.00250$ & $0.03457 \pm 0.00000$ & $0.25173 \pm 0.00018$ \\
\textbf{UNet mMIP} & $0.75771 \pm 0.00816$ & $0.03397 \pm 0.00002$ & $0.26898 \pm 0.00029$ \\
\textbf{UNet MSS mMIP} & $\textbf{0.79332} \pmb{\pm} \textbf{0.00385}$ & $\textbf{0.03564} \pmb{\pm} \textbf{0.00001}$ & $\textbf{0.24759} \pmb{\pm} \textbf{0.00022}$ \\ \bottomrule
\end{tabular}%
        }
         \caption{Quantitative evaluation of underlying vasculature}
        \label{tab:VasculatureWAA}
    \end{subtable}
\caption{Metrics comparisons of 5 segmentation results of test volume with wrap-around artefact over 5-fold cross-validation.}
\label{tab:Metrics Comparison of WAA}
\end{table}

The presence of wrap-around artefacts observed in one volume, as shown in Fig. \ref{fig:wrap-around}, results in over-segmentation of the vessels on the skull and disrupts vessel boundaries, thus disregarding the continuity of the vessels in the vicinity of the artefacts. The authors evaluated the models separately only on this volume across five folds to assess the robustness of these models against such artefacts.  

Comparison of the overall segmentation scores tabulated in Table~\ref{tab:OverallScoreWAA} shows that the proposed UNet MSS mMIP method outperforms the other baselines and the proposed methods in the presence of these artefacts, with a median Dice of $65.946 \pm 0.084$ across five cross-validation folds. A similar trend is observed in the quantitative comparison of the vasculature, as shown in Table~\ref{tab:VasculatureWAA}. 

Incorporation of MIP loss is seen to improve overall segmentation scores with UNet and UNet MSS baselines. Table~\ref{tab:VasculatureWAA} shows a considerable improvement in the identification of the underlying vasculature when these baselines are trained with the proposed single- and multi-axes MIP loss. It is also observable that the multi-axes MIP loss outperforms the single-axis MIP loss counterparts in the presence of the artefacts.

The proposed UNet MSS mMIP is found to be the method with the best performance when evaluated on the volume with wrap-around artefacts. The multi-axes method also appears to outperform its single-axis counterpart both in terms of overall segmentation and the underlying vasculature.  



\subsection{Qualitative Evaluation}
\begin{figure*}
\centering
\includegraphics[width=\textwidth]{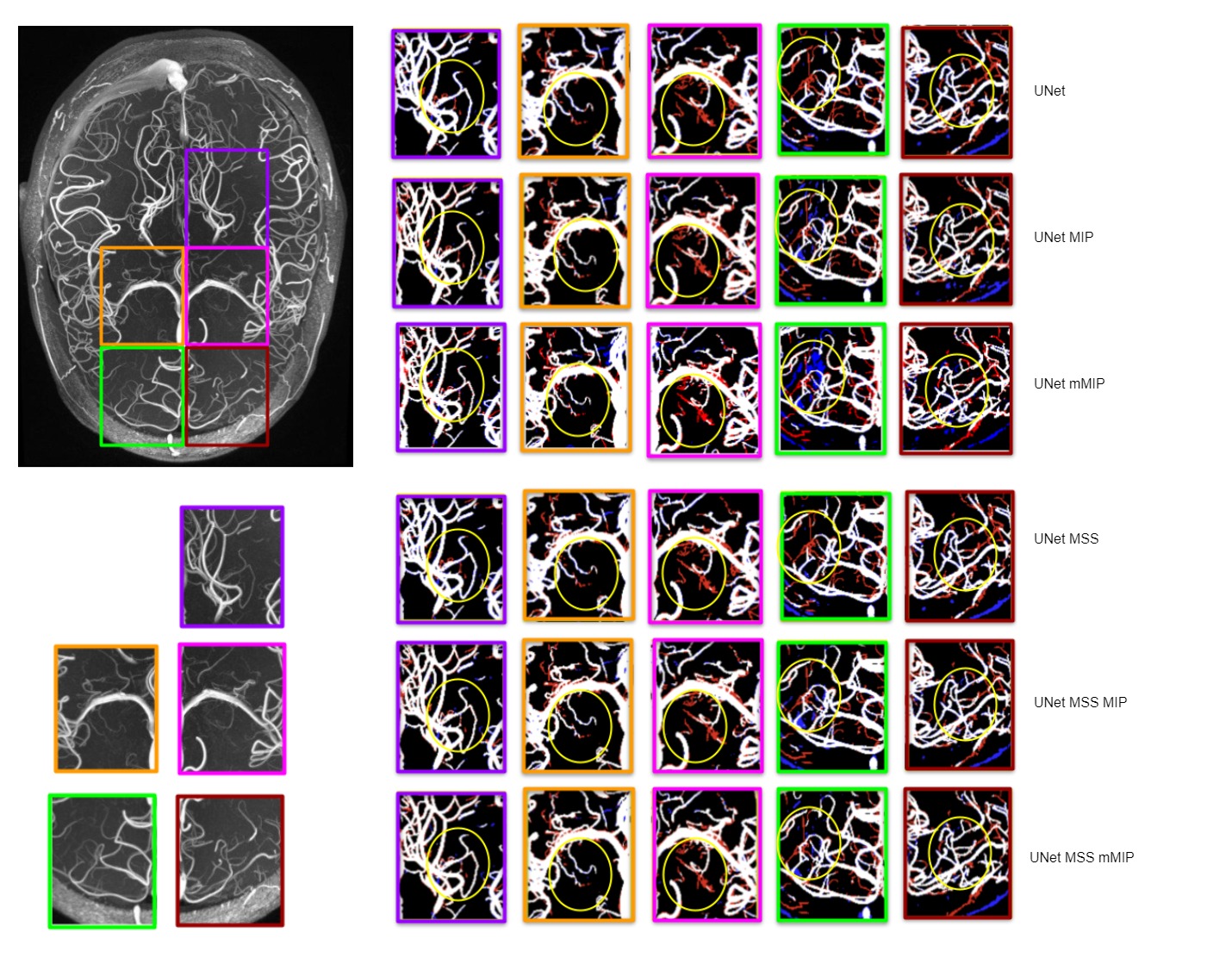} 
\caption{Visual comparison of ROIs of MIPs of segmentations resulting from training all the models with MIP loss along the z-axis and MIP loss along multiple axes. Annotations on the MIPs show the missing continuity of the vessels and the improvements after using the z-axis and multi-axes MIP loss. Each image is an overlay of network predictions on the ground truth, where blue and red represent false positives and false negatives, respectively.}
\label{fig:ROI Comparison}
\end{figure*}

\begin{figure}
\centering
\includegraphics[width=0.45\textwidth]{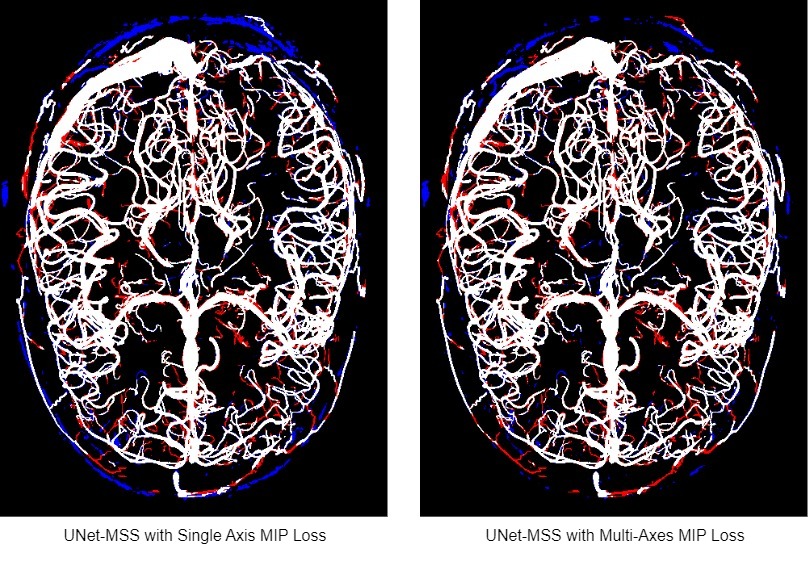} 
\caption{Comparison of False Positives between segmentations generated by UNet-MSS\_MIP and UNet MSS mMIP. Each image is an overlay of network predictions on the Ground Truth, where blue annotations represent false positives and red annotations represent false negatives.}
\label{fig:FP Comparison}
\end{figure}

The qualitative evaluation was performed by selecting five regions of interest~(ROI) exhibiting notable presence of lenticulostriate arteries, as depicted in Fig. \ref{fig:ROI Comparison}. The ROIs are marked with five different colours on the MIP of one of the test volumes, and their respective segmentations resulting from baselines and proposed approaches are also tabulated. Vessels in red and blue denote false negatives and false positives, respectively, while white denotes correctly segmented vessels. Yellow circles mark the notable differences among the different methods. The annotations on the ROIs of the segmentation results of the baselines~(UNet and UNet MSS) show the discontinuity of the vessels. Visual comparison of these ROIs against their single-axis MIP counterparts reveals improved vessel continuity. It is also evident that the models with multi-axes MIP loss prevail over the baselines. The ROI comparisons between single-axis and multi-axes MIP-based models show a reduction in false positives. This is also evident in Fig. \ref{fig:FP Comparison}, especially in the skull. ROIs show that the multi-axes MIP-based models appear to over-segment the vessels. However, a closer look at the MIP of the input volume reveals that these overextensions are still part of the vessel, which is absent in the ground truth.

\section{Conclusion}
This paper proposed a MIP-based loss term to improve vessel continuity and overall segmentation performance of deep learning models, and evaluated this loss term on two different deep learning models. It was demonstrated that the proposed MIP-based methods outperform the baselines, both quantitatively and qualitatively. The generated segmentations not only show improvement in the continuity of the vessels, but also identify structures of small vessels that were missed in the resulting segmentations from the baselines. It was observed from experiments that the voxel similarity loss (MSS loss) retains a higher importance compared to the MIP loss, as learning suffers if voxel intensity comparisons inherent in MSS loss are insufficient. Between the two types of MIP losses explored here - single- and multi-axes - the single-axis is more stable and widely applicable to different models, while the multi-axes might be advantageous for some and over-regularise other models. Overall, the UNet model with multi-axes MIP outperformed all other models (including the baselines), resulting in a Dice score of $80.245 ± 0.129$ compared. 

In future work, the authors propose the exploration of volumetric MIP loss replacing the patch-based MIP loss proposed here for training the model. The hypothesis is that this would allow the network to perceive complex vessel structures that are missing in the partial information available in a patch. Furthermore, the merits of incorporating MIP information in semi-supervised learning approaches, such as deformation-aware learning~\cite{Chatterjee_2022}, can also be a future direction for exploration. 

\bibliography{mybibfile}

\end{document}